\documentclass{aa} 
\bibpunct{(}{)}{;}{a}{}{,} 
\usepackage{txfonts}
\usepackage{natbib}

\usepackage{graphicx}
\usepackage{graphics}
\usepackage{latexsym}
\usepackage{amsfonts,amsmath,amssymb}
\usepackage{url}
\usepackage[utf8]{inputenc}
\usepackage{fancyref}
\usepackage{hyperref}
\usepackage{float}
\hypersetup{colorlinks=false,pdfborder={0 0 0}}

\begin{document}

\title{M dwarfs in the b201 tile of the VVV survey}
\subtitle{Colour-based Selection, Spectral Types and Light Curves}

\author{B{\'{a}}rbara Rojas-Ayala\inst{\ref{inst1}} \and Daniela Iglesias\inst{\ref{inst2}} \and Dante Minniti\inst{\ref{inst3}, \ref{inst4}, \ref{inst5}, \ref{inst6}} \and Roberto K. Saito\inst{\ref{inst7}} \and Francisco Surot\inst{\ref{inst3}}}

\institute{  Instituto de Astrof\'{\i}sica e Ci\^{e}ncias do Espa\c{c}o, Universidade do Porto, CAUP, Rua das Estrelas, PT4150-762 Porto, Portugal \email{babs@astro.up.pt}\label{inst1}
\and Departamento de F\'{\i}sica y Astronom\'{\i}a, Facultad de Ciencias, Universidad de Valpara\'{\i}so, Avenida Gran Breta\~{n}a 1111, Valpara\'{\i}so 2360102, Chile\label{inst2}
\and Departamento de Astronom\'{\i}a y Astrof\'{\i}sica, Pontificia Universidad Cat\'{o}lica de Chile, Vicu\~{n}a Mackenna 4860, Casilla 306, Santiago 22, Chile\label{inst3}
\and Millennium Institute of Astrophysics, Av. Vicu\~{n}a Mackenna 4860, 782-0436 Macul, Santiago, Chile\label{inst4} 
\and Vatican Observatory, Vatican City State V-00120, Italy\label{inst5}  
\and Departamento de Ciencias F\'{\i}sicas, Universidad Andr\'{e}s Bello, Rep\'{u}blica 220, Santiago, Chile\label{inst6}
\and Universidade Federal de Sergipe, Departamento de F\'{\i}sica, Av. Marechal Rondon s/n, 49100-000, S\~{a}o Crist\'{o}v\~{a}o, SE, Brazil\label{inst7}
}

\abstract 
{The intrinsically faint M dwarfs are the most numerous stars in the Galaxy, have main-sequence lifetimes longer than the Hubble time, and host some of the most interesting planetary systems known to date. Their identification and classification throughout the Galaxy is crucial to unravel the processes involved in the formation of planets, stars and the Milky Way. The ESO Public Survey VVV is a deep near-IR survey mapping the Galactic bulge and southern plane. The VVV b201 tile, located in the border of the bulge, was specifically selected for the characterisation of M dwarfs.} 
{We used VISTA photometry to identify M dwarfs in the VVV b201 tile, to estimate their subtypes, and to search for transit-like light curves from the first 26 epochs of the survey}  
{UKIDSS photometry from SDSS spectroscopically identified M dwarfs was used to calculate their expected colours in the $YJHK_s$ VISTA system. A colour-based spectral subtype calibration was computed. Possible giants were identified by a $(J-K_s, H_{J})$ reduced proper motion diagram. The light curves of 12.8<$K_s$<15.8 colour-selected M dwarfs were inspected for signals consistent with transiting objects.} 
{We identified 23,345 objects in VVV b201 with colours consistent with M dwarfs. We provided their spectral types and photometric distances, up to $\sim$ 300 pc for M9s and $\sim$ 1.2 kpc for M4s, from photometry. In the range 12<$K_s$<16, we identified 753 stars as possible giants out of 9,232 M dwarf candidates. While only the first 26 epochs of VVV were available, and 1 epoch was excluded, we were already able to identify transit-like signals in the light curves of 95 M dwarfs and of 12 possible giants.} 
{Thanks to its deeper photometry ($\sim$4 magnitudes deeper than 2MASS), the VVV survey will be a major contributor to the discovery and study of M dwarfs and possible companions towards the center of the Milky Way.}

\maketitle

\section{Introduction}

Stars with masses less than 0.6 M$_\odot$ span the peak of the stellar initial mass function and dominate the galactic stellar populations by number \citep{2010ARA&A..48..339B}. These objects are the M dwarfs: cool and faint stars, with complex spectra characterised by molecular absorption of TiO, CaH and VO in the optical, and FeH and H$_2$O in the near infrared \citep{1943QB881.M6......., 1976A&A....48..443M}. Their main sequence lifetimes are longer than the age of the universe, with the least massive (M$_\star$$<$0.25 M$_\odot$) remaining fully convective during their evolution \citep{1997ApJ...482..420L}. Some of them exhibit strong magnetic fields that can produce more magnetic activity that the sun \citep{1996ApJ...459L..95J}. They are the hosts of the closest rocky planets to the Earth, and overall, they should be the most likely hosts of terrestrial planets in the Galaxy \citep{2013A&A...549A.109B, 2013ApJ...767...95D,2013ApJ...767L...8K,2014arXiv1403.0430T}.
 
In the past decade, the study of M dwarfs has greatly benefited from photometric optical and near-infrared wide field deep surveys, such as the Sloan Digital Sky Survey \citep[SDSS,][]{2000AJ....120.1579Y}, the Two Micron All Sky Survey \citep[2MASS,][]{2006AJ....131.1163S} and the UKIRT Infrared Deep Sky Survey \citep[UKIDSS,][]{2007MNRAS.379.1599L}. Such surveys have found nearby new low-mass and ultra cool dwarfs by colour-selection and proper motion searches \citep[e.g.][]{1997ApJ...476..311K, 2009MNRAS.394..857D, 2012A&A...542A.105L}, have provided fundamental properties of a large number of low-mass stars from colour-based relations \citep[e.g. parallaxes and spectral types, ][]{2002AJ....123.3409H}, have enlightened their magnetic activity \citep[e.g. ][]{2004AJ....128..426W, 2012AJ....144...93M} and flaring properties \citep[e.g.][]{1999ApJ...519..345L, 2010AJ....140.1402H, 2012ApJ...748...58D}, and have allowed the measurements of mass and luminosity functions of low-mass dwarfs in the Galactic disk \citep{2008AJ....136.1778C, 2010AJ....139.2679B}, as well as the photometric initial mass function from Galactic clusters \citep[e.g.][]{2012MNRAS.422.1495L, 2012MNRAS.426.3403L, 2012MNRAS.426.3419B, 2013MNRAS.431.3222L}.

Of the mentioned surveys, only 2MASS mapped the bulge the Milky Way down to magnitude $\sim$14, in two epochs.  VISTA Variables in the V\'ia L\'actea (VVV) is a public ESO near- infrared (near-IR) variability survey aimed at scanning the Milky Way Bulge and an adjacent section of the mid-plane \citep{2010NewA...15..433M}. VVV complements previous near-IR surveys, providing better spatial resolution and deeper photometry ($\sim$ 4 magnitudes deeper than 2MASS) and multi-epoch $K_{s}$-band images which allows the identification of nearby faint/late M dwarfs as well as faraway unknown early M dwarfs with variable photometry consistent with transiting companions \citep{2011RMxAC..40..221S}. 

We present a colour-based selection of M dwarfs in the b201 tile of the VVV survey. In section 2, we give the description of the survey and of the tile b201. In section 3, we present our M dwarf selection method based on 6 colour-selection cuts obtained from SDSS spectroscopically observed M dwarfs with UKIDSS photometry. A spectral subtype calibration based on $(Y-J)$, $(Y-K_s)$, and $(H-K_s)$ is given in section 4. In section 5, we identify possible giants contaminants from a reduced proper motion criterion. In section 6, we identify M dwarf candidates with transit-like light curves. We discuss our results and conclusions in section 7.

\section{Data}
\begin{figure}
	\resizebox{\hsize}{!}{\includegraphics{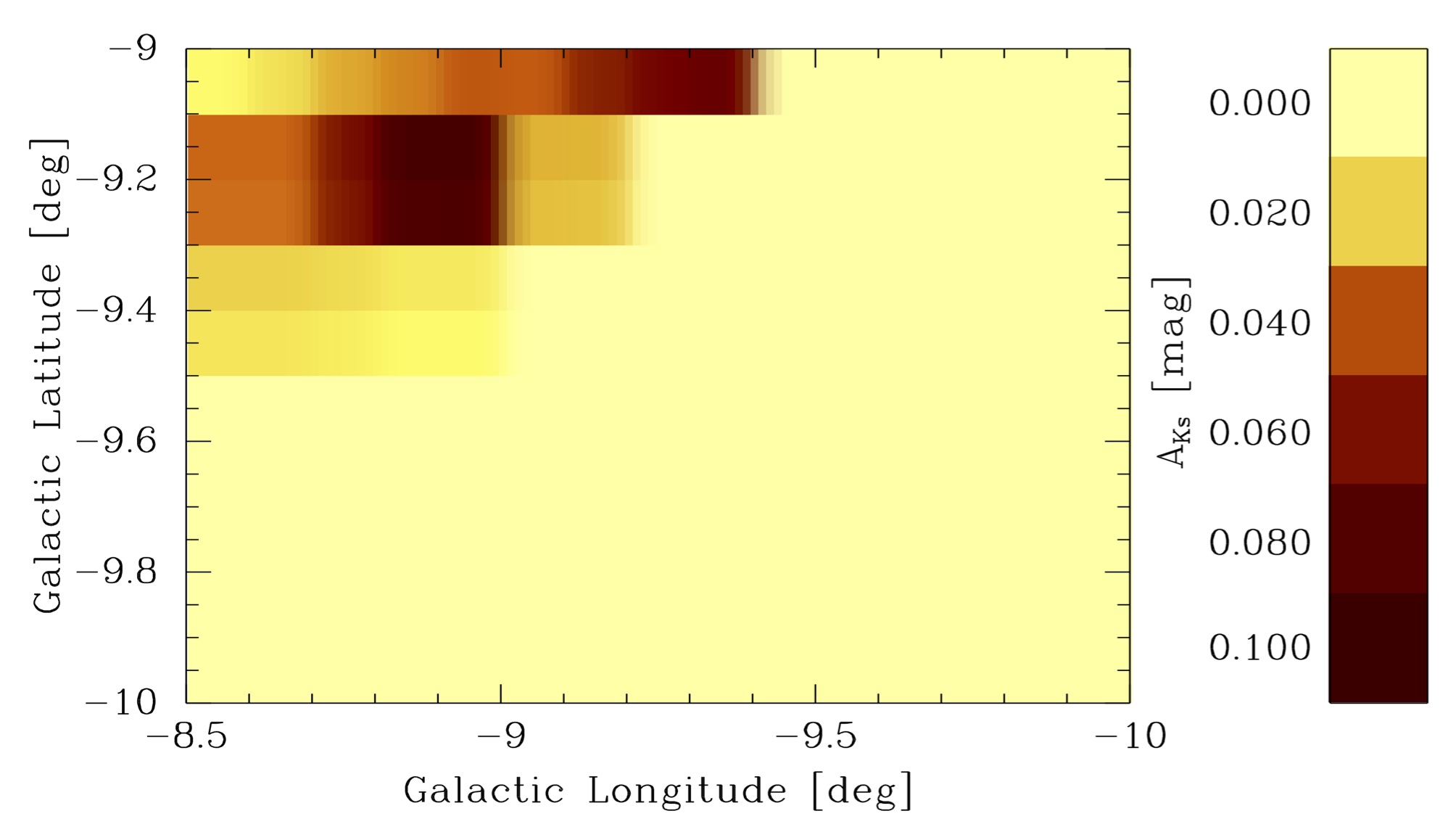}}
	\caption{Extinction map of the VVV field b201. The values are based on the maps of \cite{2012A&A...543A..13G}, using the \cite{1989ApJ...345..245C} extinction law, for a resolution of $6\arcmin \times 6\arcmin$. This field is located in the bottom-right corner of the VVV bulge area, within coordinates $-10^{\circ} \lesssim l \lesssim -8.5^{\circ}$ and $-10^{\circ} \lesssim l \lesssim -9^{\circ}$. For a large portion of the area the total extinction $A_{\rm K_s}$ is lesser than the lower limit computed by the \cite{2012A&A...543A..13G} map, of $A_{\rm K_s}<0.0001$~mag. While the maximum value found is $A_{\rm K_s}=0.0890$~mag, the mean extinction over the entire field is $A_{\rm K_s}=0.0083$~mag. The total extinction is expected to be mostly in the background of the M dwarfs, and therefore overestimated. However, the values are small and the effect should be negligible, at least in this particular region (the effect would be significant in other more reddened regions of the bulge as shown by the reddening maps of \cite{2012A&A...543A..13G}). This degree of extinction produces no or negligible effects in the photometric limits used in our target selection.}
	\label{extinction}
\end{figure}

The VVV survey gives near-IR multi-colour information in five passbands: $Z$ (0.87 $\mu m$), $Y$ (1.02 $\mu m$), $J$ (1.25 $\mu m$), $H$ (1.64 $\mu m$), and $K_s$ (2.14 $\mu m$) which complements surveys such as 2MASS \citep{2006AJ....131.1163S}, DENIS \citep{1997Msngr..87...27E}, GLIMPSE II \citep{2005sptz.prop20201C}, VPHAS+\citep{2013Msngr.154...41D}, MACHO \citep{1993ASPC...43..291A}, OGLE \citep{1992AcA....42..253U}, EROS \citep{1993Msngr..72...20A}, MOA \citep{1999PThPS.133..233M}, and GAIA \citep{2001A&A...369..339P}. The survey covers a 562 square degree area in the Galactic bulge and the southern disk which contains \textasciitilde{}$10^{9}$ point sources \citep{2012A&A...537A.107S}. Each unit of VISTA observations is called a (filled) \textquotedblleft{}tile\textquotedblright{}, consisting of six individual (unfilled) pointings (or \textquotedblleft{}pawprints\textquotedblright{}) and covers a 1.64 $deg^{2}$ field of view. To fill up the VVV area, a total of 348 tiles are used, with 196 tiles covering the bulge (a 14 $\times$ 14 grid) and 152 for the Galactic plane (a 4 $\times$ 38 grid) \citep{2012A&A...544A.147S}. We selected one specific tile from the bulge to characterise M dwarf stars called ``b201''. The galactic coordinates of this tile's center are $l$=350.74816 and $b$=-9.68974. This tile is located in the border of the bulge where star density is lower and extinction is small allowing good photometry, as shown in Figure \ref{extinction}. Photometric catalogues for the VVV images are provided by the Cambridge Astronomical Survey Unit (CASU\footnote{http://apm49.ast.cam.ac.uk/}). The catalogues contain the positions, fluxes, and some shape measurements obtained from different apertures, with a flag indicating the most probable morphological classification. In particular, we note that \textquotedblleft{}-1\textquotedblright{} is used to denote the best-quality photometry of stellar objects \citep{2012A&A...544A.147S}. Some other flags are \textquotedblleft{}-2\textquotedblright{} (borderline stellar), \textquotedblleft{}0\textquotedblright{} (noise), \textquotedblleft{}-7\textquotedblright{} (sources containing bad pixels), and \textquotedblleft{}-9\textquotedblright{} (saturated sources).

\begin{figure*}
	\resizebox{\hsize}{!}{\includegraphics{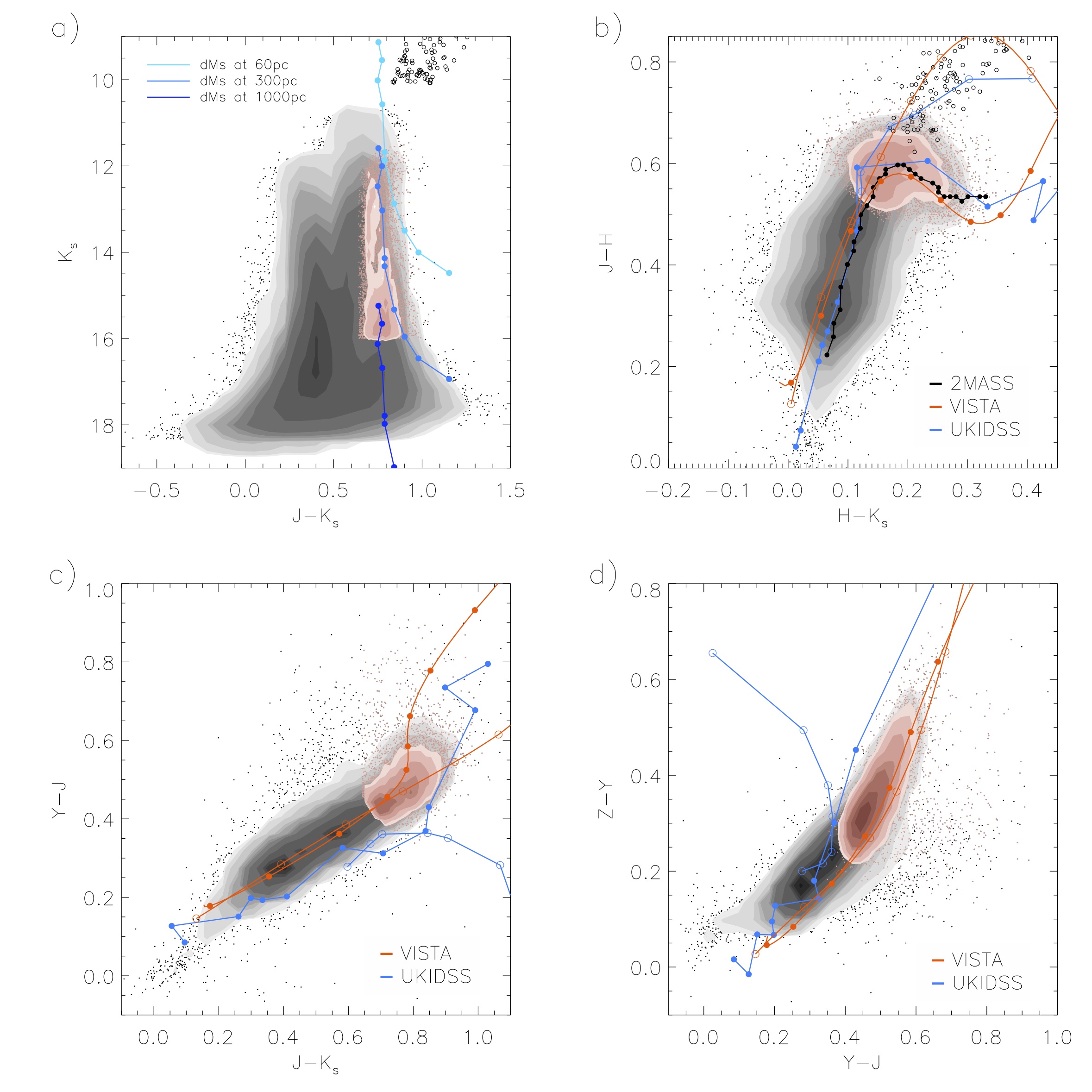}}
	\caption{Colour-Magnitude diagram and colour-colour diagrams for objects classified as "stellar" in the tile b201. The colour-based selection of M dwarfs is shown in pink in all diagrams. Black open circles are giant stars located at similar $(l,b)$ as b201 by the BRaVA Project \citep[]{2007ApJ...658L..29R,2012AJ....143...57K}. a) The colour identified M dwarfs fill the region that agrees with the K+M dwarf sequence in the $(J-K_s,K_s)$ CMD of the outermost region in the VVV bulge area, identified by \citet[]{2012A&A...544A.147S}. The estimated M dwarf spectral sequences at 60 pc, 300 pc and 1000 pc are shown. b), c), and d) $(H-K_s,J-H)$, $(J-K_s,Y-J)$, and $(Y-J,Z-Y)$ diagrams for all "stellar" objects with 12<$K_s$<16 only. The blue lines represent the dwarf (filled circles) and giant (open circles) sequences, based on UKIDSS synthetic colours by \citet[]{2006MNRAS.367..454H}.The orange lines represent the dwarf (filled circles) and giant (open circles) sequences, based on VISTA synthetic colours derived from stars in the IRTF Spectral Library (see APPENDIX \ref{appendixa}). In the $(H-K_s,J-H)$ diagram, the black line corresponds to the $JHK_s$ stellar locus by \citet[]{2007AJ....134.2398C}, derived from SDSS and 2MASS photometry), and it is in agreement with the one derived in this work from VISTA synthetic colours. The disagreement between the UKIDSS and VISTA sequences may be due to the differences between the $Z$ and $Y$ synthetic filters of each survey.}
	\label{colours_diag}
\end{figure*}

\section{Colour-Selection Cuts from SDSS-UKIDSS M dwarfs}

To identify potential M dwarfs in the VVV b201 tile, we performed several colour-selection cuts using the VVV passbands as described in the subsections below. Before performing those cuts, we did a pre-selection of the objects in the tile b201 to ensure that the objects have the best-quality photometry. In this pre-selection, we included only objects that had photometry in all five passbands ($ZYJHK_s$), and that were classified as "stellar" in each passband. A total of 142,321 objects in the tile b201 satisfied these conditions.  

The colour selection cuts were defined from spectroscopically identified M dwarfs with UKIRT Infrared Deep Sky Survey (UKIDSS) photometry. 

We selected the Sloan Digital Sky Survey DR7 Spectroscopic M dwarf catalogue by \citet{2011AJ....141...97W} as the comparative M dwarf sample. The 70,841 M dwarf stars in this catalogue had their optical spectra visually inspected and the spectral type of each object was assigned by comparing them to spectral templates. Their spectral types range from M0 to M9, with no half subtypes. This catalogue also provides values for the CaH2, CaH3 and TiO5 indices, which measure the strength of CaH and TiO molecular features present in the optical spectra of M dwarfs.

\begin{table*} 
\caption{VISTA mean colours and standard deviations per spectral subtype.}
\begin{tabular}{l l l l l l l l}
\hline\hline
Sp.T. & $\overline{Y-J}$ & $\overline{Y-H}$ & $\overline{Y-K_s}$ &  $\overline{J-H}$ & $\overline{J-K_s}$  & $\overline{H-K_s}$  & $\#$ stars  \\ 
\hline
  M0 & 0.428 $\pm$ 0.092 & 1.039 $\pm$ 0.087 & 1.163 $\pm$ 0.063 & 0.611 $\pm$ 0.116 & 0.734 $\pm$ 0.092 & 0.124 $\pm$ 0.079 & 1946\\
  M1 & 0.449 $\pm$ 0.077 & 1.047 $\pm$ 0.061 & 1.200 $\pm$ 0.064 & 0.598 $\pm$ 0.086 & 0.751 $\pm$ 0.081 & 0.153 $\pm$ 0.046 & 2520\\
  M2 & 0.467 $\pm$ 0.061 & 1.042 $\pm$ 0.073 & 1.219 $\pm$ 0.058 & 0.575 $\pm$ 0.088 & 0.752 $\pm$ 0.071 & 0.177 $\pm$ 0.058 & 3043\\
  M3 & 0.487 $\pm$ 0.081 & 1.043 $\pm$ 0.062 & 1.241 $\pm$ 0.064 & 0.556 $\pm$ 0.089 & 0.754 $\pm$ 0.083 & 0.198 $\pm$ 0.038 & 3293\\
  M4 & 0.515 $\pm$ 0.083 & 1.057 $\pm$ 0.090 & 1.278 $\pm$ 0.068 & 0.542 $\pm$ 0.110 & 0.762 $\pm$ 0.085 & 0.220 $\pm$ 0.075 & 2872\\
  M5 & 0.555 $\pm$ 0.096 & 1.092 $\pm$ 0.069 & 1.340 $\pm$ 0.082 & 0.538 $\pm$ 0.103 & 0.786 $\pm$ 0.099 & 0.248 $\pm$ 0.044 & 1264\\
  M6 & 0.619 $\pm$ 0.082 & 1.150 $\pm$ 0.067 & 1.442 $\pm$ 0.076 & 0.531 $\pm$ 0.087 & 0.823 $\pm$ 0.084 & 0.292 $\pm$ 0.033 & 1224\\
  M7 & 0.664 $\pm$ 0.117 & 1.198 $\pm$ 0.126 & 1.513 $\pm$ 0.136 & 0.533 $\pm$ 0.064 & 0.849 $\pm$ 0.068 & 0.315 $\pm$ 0.037 & 1141\\
  M8 & 0.758 $\pm$ 0.070 & 1.304 $\pm$ 0.102 & 1.662 $\pm$ 0.122 & 0.546 $\pm$ 0.052 & 0.904 $\pm$ 0.067 & 0.358 $\pm$ 0.033 & 320\\
  M9 & 0.850 $\pm$ 0.079 & 1.429 $\pm$ 0.114 & 1.830 $\pm$ 0.139 & 0.579 $\pm$ 0.054 & 0.980 $\pm$ 0.071 & 0.401 $\pm$ 0.038 & 151\\
\hline
\end{tabular}
\label{spec_color}
\end{table*}

We performed a cone search with a radius of 5$\arcsec$ of these SDSS M dwarf stars in the UKIDSS-DR8 survey \citep{2012yCat.2314....0L}. The UKIDSS survey is carried out using the Wide Field Camera (WFCAM), with a $Y$ (1.0um), $J$ (1.2um), $H$ (1.6um) and $K$ (2.2um) filter set. The cone search provided UKIDSS-DR8 matches for almost half of the SDSS M dwarf sample (34,416 matches) . Next, we only kept the UKIDSS counterparts consistent with being a stellar objects (pStar > 0.9), with measured magnitudes in all WFCAM $YJHK$ filters, and with CaH and TiO indices compatible with average M dwarf stars. The final SDSS-UKIDSS comparative M dwarf sample consists of 17,774 objects.

To convert the WFCAM $YJHK$ magnitudes of the SDSS-UKIDSS M dwarf sample to VISTA $YJHK_s$ magnitudes, we used the conversions provided by CASU \footnote{http://apm49.ast.cam.ac.uk/surveys-projects/vista/technical/photometric-properties}, derived from regions observed with both VISTA and WFCAM.

The mean and standard deviation for all of the colours from VISTA $YJHK_s$ photometry per M spectral subtype, as well as the number of stars considered for their computation, are shown in Table \ref{spec_color}.

We have defined the colour-based selection of  M dwarf by inspecting the colours of all the stars within 1-sigma from the mean colour. The
resulting limits are:

\begin{description}
\item 0.336 < $(Y-J)_{VISTA} $< 0.929
\item 0.952 < $(Y-H)_{VISTA} $< 1.544
\item 1.100 < $(Y-K_s)_{VISTA}$ < 1.969
\item 0.432 < $(J-H)_{VISTA} $< 0.727
\item 0.642 < $(J-K_s)_{VISTA}$ < 1.051
\item 0.045 < $(H-K_s)_{VISTA}$ < 0.438.
\end{description}

From our pre-selection of 142,321 objects, only 23,345 objects have colours that are consistent with M dwarf stars, according to the colour-cuts shown above. Forty-percent of these objects have magnitudes 12$<$K$_s$$<$16,  and therefore have reliable magnitudes for variability and are the best M dwarf candidates to detect any possible transits (9,232 objects). 

\section{Spectral Types and Photometric Distances for VVV M dwarfs}

The mean colours per spectral type in Table \ref{spec_color} show that spectral type is a monotonically increasing function for the following colours: $Y-J$, $Y-K_s$, and $H-K_s$. We conducted multivariate regressions on the $Y-J$, $Y-K_s$, and $H-K_s$ colours for the 17,774 stars in the SDSS-UKIDSS comparative M dwarf sample to identify the best-fit relationship to predict each star's spectral type. The resulting subtype calibration is

\begin{eqnarray}
\label{msubtype1}
\mathrm{M subtype} & =  5.394 \mathrm{(Y\!\!-\!\!J)} + 4.370 \mathrm{(Y\!\!-\!\!J)^2}\nonumber \\
& + 24.325 \mathrm{(Y\!\!-\!\!K_s)} - 7.614 \mathrm{(Y\!\!-\!\!K_s)^2} \nonumber  \\
& + 7.063 (\mathrm{H\!\!-\!\!K_s)} -20.779,   \\
\mathrm{RMSE_V} & =  1.109, \nonumber
\end{eqnarray}
with $RMSE_V$ being the root-mean-square error of validation, a sensible estimate of average prediction error \citep[see appendix in][]{2012ApJ...748...93R}. Spectral types for all the M dwarf candidates are given in Table \ref{final_list}.

To identify the location of M dwarfs at different distances in the Colour-Magnitude Diagram (CMD), we used the nearby M dwarfs with $M_{K_s}$ and spectral type estimates in \citet{2012ApJ...748...93R}. Using the colour transformations from WFCAM to the VISTA system, we estimated the apparent $K_s$ magnitudes at different distances per spectral type (Table \ref{absolutemag}). The locations of the M dwarf sequence at 60 $\mathrm{pc}$, 300 $\mathrm{pc}$ and 1000 $\mathrm{pc}$ coincide with the location of the colour-based selection of M dwarfs described above, as well as the K+M dwarf sequence identified by \citet{2012A&A...544A.147S}, as shown in the CMD of Figure \ref{colours_diag}. Based on the estimated M subtypes derived by Equation \ref{msubtype1}, we provide estimated distances for the colour-based selected M dwarfs in Table \ref{final_list}. We emphasise that these distances are provided to have an approximate location of the objects with respect to the bulge, and they are not expected to be accurate. 

Considering the spectral types in Table \ref{final_list}, the deeper photometry of VVV has a higher impact in the number of late type M dwarfs that can be found in the b201 tile.  By performing a 5$\arcsec$ search of the 23,345 objects in 2MASS Point Source catalogue \citep{2003yCat.2246....0C}, we can estimate that the number of M9 stars found by VVV in the b201 tile is $\sim$30 times larger than the one that can be found with only 2MASS photometry (1 versus 30 M9 stars at distances up to $\sim$ 300 pc).  The number of M8 and M7 stars is 18 and 13 times larger that the ones by only 2MASS photometry (at distances up to $\sim$ 500 pc), while the number of M4 is about 4 times more (at distances up to $\sim$ 1000 pc). 

\begin{table}
\caption{Average $M_{K_s}$ and $K_s$ at different distances per M subtype.}
\begin{tabular}{l l l l l l} 
\hline\hline
Sp.T. & $M_{K_s}$ &  $(J-K_s)$ & $K_s^{60\mathrm{pc}}$ & $K_s^{300\mathrm{pc}}$ & $K_s^{1000\mathrm{pc}}$ \\ 
\hline
  M0 & 5.240 & 0.753 & 9.131 & 12.626 & 15.240\\
  M1 & 5.656 & 0.773 & 9.547 & 13.042 & 15.656\\
  M2 & 6.126 & 0.748 & 10.017 & 13.512 & 16.126\\
  M3 & 6.681 & 0.775 & 10.572 & 14.067 & 16.681\\
  M4 & 7.790 & 0.788 & 11.680 & 15.175 & 17.790\\
  M5 & 7.976 & 0.788 & 11.866 & 15.361 & 17.976\\
  M6 & 8.980 & 0.842 & 12.871 & 16.366 & 18.980\\
  M7 & 9.609 & 0.901 & 13.499 & 16.994 & 19.609\\
  M8 & 10.113 & 0.980 & 14.003 & 17.498 & 20.113\\
  M9 & 10.589 & 1.153 & 14.480 & 17.975 & 20.589\\
\hline
\end{tabular}
\label{absolutemag}
\end{table}

\section{Possible Giant Contaminants}

Giant stars are the most common contaminants of colour-based selections of M dwarfs. \citet{1988PASP..100.1134B} derived intrinsic colours in the Johnson-Glass system for several V and III class stars, and schematically showed the position of dwarfs, giants, supergiants, carbon stars and long-period variables in the $(H-K,J-H)$ diagram. By using $(V-K)$ as proxy for spectral type, \citet{1988PASP..100.1134B}  showed that giants and dwarf stars share similar $(J-H)$ and $(H-K)$ colours for $(V-K)$< 3.5, but their $(H-K)$ colours make them almost indistinguishable up to $(V-K)$$\sim$6. Colour cuts based on $(J-K_s)$ and $(J-H)$ colours have been used to identify giants in different parts of the Galaxy \citep[e.g.][]{2010ApJ...722..750S, 2014AJ....147...76B}, however they only serve to isolate the cooler giants from M dwarfs ($(J-K_s)$> 0.85). The giant sequence passes through the M dwarf region in the $(J-K_s, J-H)$ diagram, with K and early M giants contaminating the sample of colour identified M dwarfs (see Figure \ref{colours_diag}). The colour selection criteria described in \citet[][]{2010ApJ...722..750S} identifies 299 objects as giants stars in our whole M dwarf sample, 60 of them within the magnitude range 12<$K_s$<16.

Another way to identify giants, when their distances are unknown, is by their location in a reduced proper motion diagram \citep[e.g.][]{2011AJ....142..138L}. To get estimates of the proper motions of the whole M dwarf colour-based selection, we performed a cone search with a 5$\arcsec$ radius of their coordinates in the PPMXL catalogue \citep{2010AJ....139.2440R} and the SPM4 catalogue \citep{2011AJ....142...15G}. The PPMXL catalogue covers both hemispheres, while the SPM4 catalogue covers objects between the south celestial pole and -20$^{\circ}$ declination, with higher proper motion precision than PPMXL. Both of these catalogues provide the crossmatched 2MASS photometry for their objects, and we only considered the objects with $\lvert K_s^{VISTA} - K_s^{2MASS} \rvert$ $\leq$ 0.5 mag. We obtained total proper motions, $\mu$ in $\arcsec$/yr,  for 6,464 and 2,940 objects from PPMXL and SPM4, respectively. The number of stars in the 12<$K_s$<16 M dwarf selection with PPMXL and SPM4 total proper motions  is 6,216 and 2,760, respectively. We calculated their $J$ magnitude reduced proper motion $H_J$ using the definition 

\begin{equation}
\label{rpm1}
H_J  =  J + 5 \log_{10}\mu. 
\end{equation}

\citet{2011AJ....142..138L} defined a criterion to separate M dwarfs from giants based on $V$ magnitude, reduced proper motion $H_V$, and $(V-J)$ colour. This criterion cannot be used for our stars since $V$ magnitudes are hard to find in the literature for the 12 < $K_s$ < 16 M dwarfs. For that reason, we computed an equivalent criterion based on $J$ and $K_s$ magnitudes. We grouped the stars of the \citet{2011AJ....142..138L} study by their estimated spectral types, obtained their mean $(V-J)$ and $(J-K_s)$ colours, and calculated the dwarf/giant discriminator $H_J^{*}$ as function of mean $(V-J)$ per spectral type, using our definition of $H_J$, by rewriting Equation 8 of \citet{2011AJ....142..138L} as follows:

\begin{equation}
\label{rpm2}
H_J^{*}  = 1.5 (V-J) - 3.0.
\end{equation}
Then, using the mean $(J-K_s)$ colour corresponding to each mean $(V-J)$ per spectral type, we performed a linear fit to obtain $H_J^{*}$ as function of $(J-K_s)$, and, therefore, an equivalent criterion to Equation \ref{rpm2} based on $(J-K_s)$, instead of $(V-J)$

\begin{equation}
\label{rpm3}
H_J^{dwarf} > H_J^{*} = 68.5 (J-K_s) - 50.7.
\end{equation}

\begin{figure*}
	\resizebox{\hsize}{!}{\includegraphics{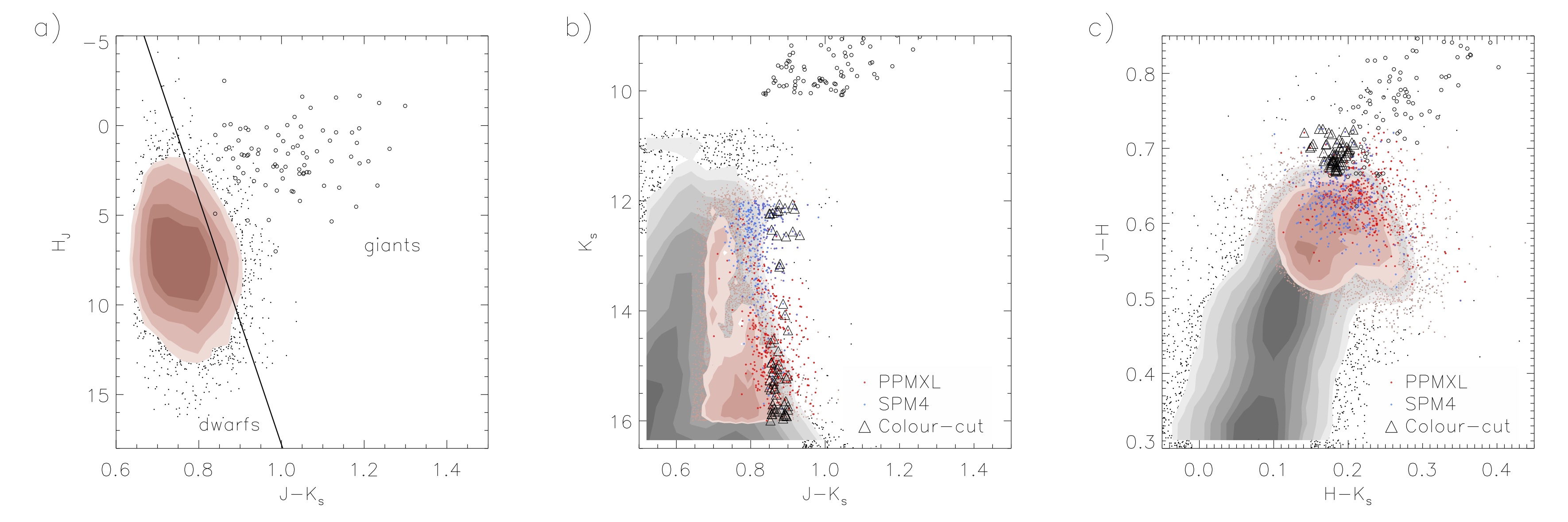}}
	\caption{Reduced proper motion (RPM), colour-magnitude, and $(H-K_s, J-H)$ diagrams for the 12<$K_s$<16 M dwarfs selected by colour-based relations (pink contours). Bulge giant stars from the BRaVA Project are shown as black open circles. a) In the RPM diagram, likely giants contaminants lie above the black line which represents the $H_J^{*}$ as function of $(J-K_s)$ colour. b) In the CMD, likely giant stars from SPM4 (blue dots) exhibit brighter magnitudes due to the magnitude cut of the survey, while PPMXL provides parallaxes for fainter likely giants (red dots).  Crosses depict likely giants identified by colour-cuts from \citet[][]{2010ApJ...722..750S}. c) Likely giants selected by colour-cuts (crosses) follow the trend of the BRaVa giants (open circles) in the $(H-K_s, J-H)$ diagram. Likely giants identified by proper motions are mostly located in the upper part of the M dwarf region (redder $J-H$ colours), and exhibit $H-K_s$ colours redder than $\sim$ 0.13 mag.}
	\label{fig_rpm}
\end{figure*}

In the 12<$K_s$<16 M dwarf sample, using the criterion above, we identified 555 likely giant stars from PPMXL proper motions, with 24 of them exhibiting $(J-K_s)$ and $(J-H)$ colours compatible with giants. From SPM4 proper motions, we identified 328 likely giants, with 18 of them exhibiting cool giant colours. Almost all of the 12<$K_s$<16 objects in SPM4 have also PPMXL proper motions (2,595 stars). For about 40$\%$ of them, the PPMXL and SPM4 total proper motions agree within $\pm$0.01 $\arcsec$/yr, with the PPMXL catalogue providing higher values of total proper motions than SPM4 (by more than 0.01$\arcsec$/yr) for the majority of the rest. Considering the reduced proper motion criterion, only 164 objects are likely giants with both PPMXL and SPM4 proper motions (16 of them have giants colours, too). Their locations in the CMD and $(H-K_s, J-H)$ diagram are shown in Figure \ref{fig_rpm}. 

The names, VISTA photometry, spectral type, and estimated distances of the 23,345 colour-selected M dwarf candidates are listed in Table \ref{final_list}. The total proper motion, $J_{RPM}$ and likely giant flag are given for the stars with PPMXL and SPM4 proper motions, as well as the 299 colour selected giants.

\begin{table*}
\caption{M dwarfs in the VVV b201 tile. Only a portion of this table is shown here to demonstrate its form and content.}
\tiny
\begin{tabular}{l l l l l l l l l l l l l l l}
\hline\hline
Name & Z & Y & J & H & K$_s$ & Sp.T. & d & PM$_{SPM4}$ & H$_{J}^{SPM4}$ & PM$_{PPMXL}$ & H$_{J}^{PPMXL}$ & Obs. \\
   & mag & mag & mag & mag & mag &  & pc & $mas/yr$ & mag & $mas/yr$ & mag &  \\
\hline
  VVVJ17594599-4205194 & 18.242 & 17.676 & 17.105 & 16.516 & 16.396 & 3 & 877 &  &  &  &  & \\
  VVVJ17594603-4206345 & 19.523 & 19.004 & 18.094 & 17.467 & 17.413 & 8 & 288 &  &  &  &  & \\
  VVVJ17594628-4206516 & 18.166 & 17.964 & 17.432 & 16.786 & 16.638 & 3 & 980 &  &  &  &  & \\
  VVVJ17594859-4206457 & 17.463 & 17.103 & 16.599 & 15.976 & 15.812 & 3 & 670 &  &  &  &  & \\
  VVVJ17594908-4205074 & 16.729 & 16.466 & 16.024 & 15.399 & 15.27 & 2 & 674 &  &  & 11.987 & 6.417 & \\
  VVVJ17594944-4207152 & 14.435 & 14.135 & 13.744 & 13.12 & 12.949 & 1 & 287 &  &  &  &  & \\
  VVVJ17594955-4205529 & 13.493 & 13.21 & 12.781 & 12.173 & 12.038 & 1 & 189 & 9.173 & 2.593 & 9.571 & 2.686 & \\
  VVVJ17594971-4205192 & 15.525 & 15.065 & 14.525 & 14.0 & 13.746 & 4 & 155 &  &  & 5.345 & 3.165 & \\
  VVVJ17594996-4205371 & 16.832 & 16.539 & 16.098 & 15.478 & 15.396 & 1 & 887 & 24.537 & 8.047 &  &  & \\
  VVVJ17595034-4207015 & 18.16 & 17.711 & 17.082 & 16.637 & 16.381 & 5 & 480 &  &  &  &  & \\
  VVVJ17595079-4207372 & 16.179 & 15.856 & 15.375 & 14.654 & 14.513 & 3 & 368 &  &  & 11.873 & 5.748 & G23\\
  VVVJ17595111-4206312 & 18.64 & 18.466 & 17.919 & 17.397 & 17.153 & 4 & 746 &  &  &  &  & \\
  VVVJ17595235-4207075 & 18.284 & 17.923 & 17.315 & 16.798 & 16.557 & 5 & 520 &  &  &  &  & \\
  VVVJ17595271-4207183 & 18.685 & 18.308 & 17.8 & 17.167 & 16.871 & 4 & 655 &  &  &  &  & \\
  VVVJ17595313-4204316 & 17.844 & 17.54 & 17.076 & 16.466 & 16.345 & 2 & 1106 &  &  & 5.946 & 5.947 & \\
\hline\end{tabular}
\tablefoot{In column Obs., likely giants are flagged as \textquotedblleft{}G\textquotedblright{} followed by numbers 1, 2 and/or 3, where 1-proper motion from SPM4, 2-proper motion from  PPMXL, and 3-colour selected. Stars with transit-like curves are flagged as \textquotedblleft{}T\textquotedblright{} }
\label{final_list}
\end{table*}

\section{VVV Light Curves for M dwarfs}

\begin{figure*}[t]
	\resizebox{\hsize}{!}{\includegraphics{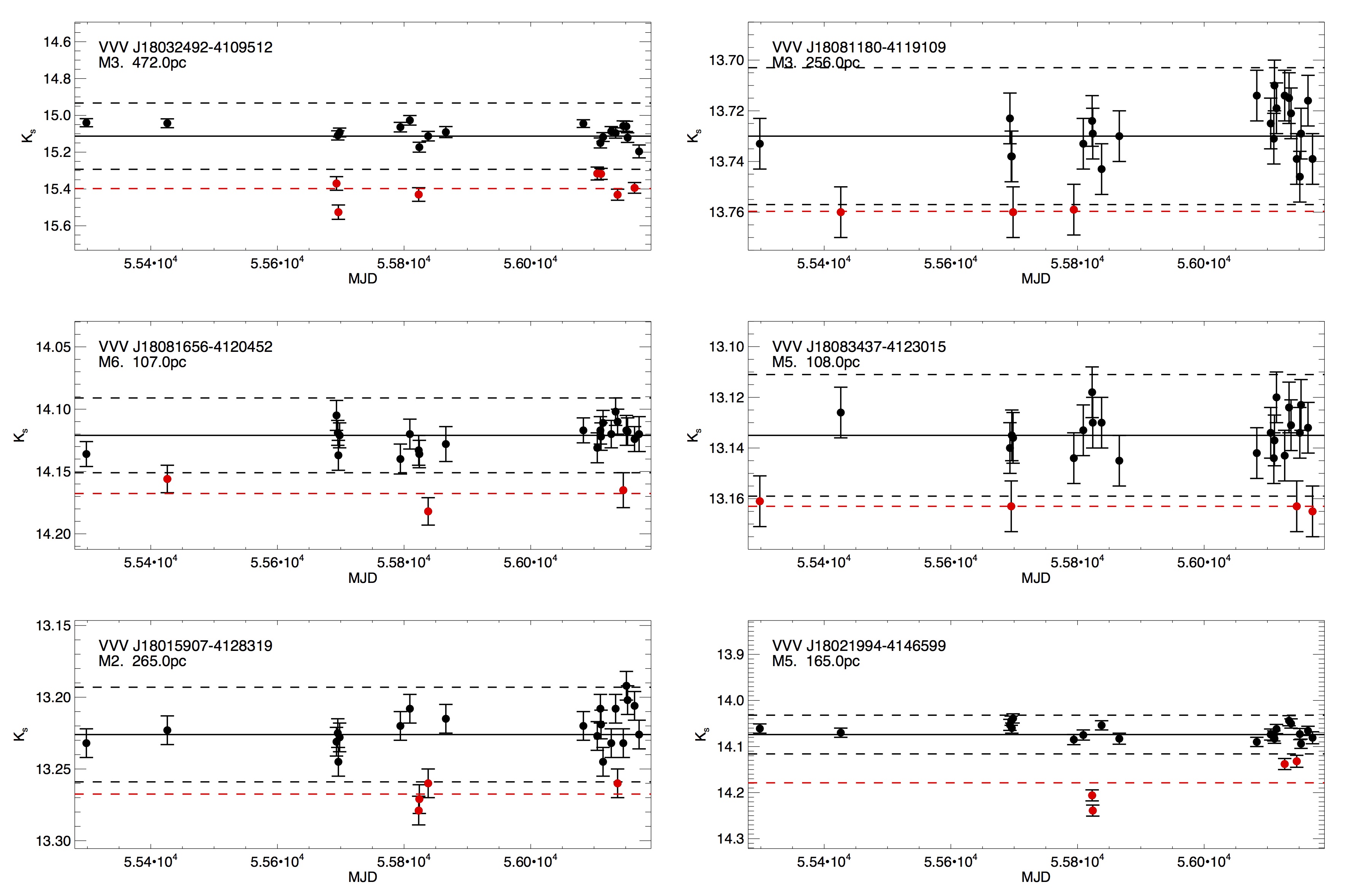}}
	\caption{Light curves of 6 M dwarf candidates. The dotted black lines indicate the $\pm$ 3-MAD cuts to find the outliers. The red dotted line indicates the mean magnitude of the outliers. Considering the mean magnitude of the outliers and assuming a non-grazing transit, the possible companions have sizes consistent with ultra-cool dwarfs down to terrestrial bodies.}
	\label{transitos}
\end{figure*}

We constructed the light curves of the 9,232 M dwarf candidates with 12<$K_s$<16  in the tile b201, considering only the first 26 epochs of VVV observations. These epochs cover observations taken from December 2010 to September 2012. Considering the unevenly sampled data and the low number of epochs, we searched for objects with light-curves with "transit-like" signals by identifying "outliers", i.e. epochs with magnitudes considerable fainter than the median magnitude of all epochs, in each light-curve. We used the median and median absolute deviation (MAD) statistics to identify outliers, given that they are very insensitive to the presence of outliers in the data, contrary to the mean and the standard deviation. An epoch is classified as an outlier if its magnitude is 3-MAD away from the median magnitude of the light-curve.

By inspecting the outliers in each light-curve, we found that certain epochs were consistently flagged as outliers. These epochs correspond to the observation dates with seeings smaller than 0.8", where several stars with $K_s$ < 12.8-13 mag appear to be consistently fainter. We also found that epoch 14 exhibits the higher dispersion in all magnitudes. Considering the above, and the fact that at all epochs the dispersion increases at $K_s$ > 15.8 mag, we restricted our study of the light-curves to the M dwarf candidates with 12.8 < $K_s$ < 15.8 mag, without considering epoch 14. Due to the small number of epochs, we also only considered objects with "-1" flag (stellar) in all epochs and with mean magnitude errors of all the epochs smaller or equal to 1.5 MAD. 

We found 95 M dwarf candidates and 12 likely giants that exhibit at least 3 epochs with 3-MAD fainter $K_s$ magnitudes than the median of the 25 epochs (removing epoch 14) out of the 3,843 objects that satisfied all of the conditions mentioned above. Examples of the type of light curves found can be seen in Figure \ref{transitos}.

\section{Conclusions}

We identified 23,345 M dwarf candidates in the VVV b201 tile from colour-cuts based on 17,774 M dwarfs with SDSS spectra and UKIDSS photometry. Their positions in VISTA-colour diagrams match the stellar locus derived from stars with 2MASS photometry by \citet{2007AJ....134.2398C}, as well as the dwarf sequences from synthetic colours calculated from IRTF Library spectra (Appendix \ref{appendixa}). From that sample, we selected 9,232 stars with 12<$K_s$<16 magnitudes for further characterisation of their light curves. From their position on a reduced proper motion diagram and their $J-K_s$ and $J-H$ colours, we identified 753 objects with a higher chance of being giants instead of M dwarfs. From the sample of likely M dwarfs, we searched for transit-like light curves based on 25 epochs of the VVV survey. We found 95 objects with light curves with 3 or more epochs with significant decrease of their luminosity. Assuming that the light curves correspond to non-grazing objects fully transiting the M dwarf, the sizes of these possible companions range from the ones of ultra-cool dwarfs ($\sim$0.12) to sizes of a couple of earth radii. However, this is assuming the conditions mentioned above and assuming that the light curves are due to transiting objects. More VVV epochs and further spectroscopic follow-up are needed to confirm the properties of the stars and the real nature of their light curves.
The deeper photometry of VVV has a higher impact in the number of nearby, late type M dwarfs that can be found towards the bulge of the Galaxy, missed by previous NIR surveys. VVV will be a major contributor to the discovery and study of very low-mass stars and and possible companions towards the center of the Milky Way.

\begin{acknowledgements} 

We would like to thank Karla Pe\~{n}a Ram\'{\i}rez and Ann M. Martin for their help during the preparation of this manuscript. We thank the anonymous referee for her/his helpful comments that improved our manuscript. We gratefully acknowledge use of data from the ESO Public Survey programme ID 179.B-2002 taken with the VISTA telescope, and data products from the Cambridge Astronomical Survey Unit, and funding from the BASAL Center for Astrophysics and Associated Technologies (CATA) PFB-06, from Project IC120009 "Millennium Institute of Astrophysics (MAS)" of Iniciativa Cient\'{i}fica Milenio from the Ministerio de Econom\'{i}a, Fomento y Turismo de Chile, and from Proyecto FONDECYT Regular 1130196.
B.R-A. acknowledges support from the Fundac\~{a}o para a Ci\^{e}cia e a Tecnologia (FCT, Portugal) through the grant contract SFRH/BPD/87776/2012.
R.K.S. acknowledges support from CNPq/Brazil through projects 310636/2013-2 and 481468/2013-7 \\

This work is based in part on data obtained as part of the UKIRT Infrared Deep Sky Survey.
This publication makes use of data products from the Two Micron All Sky Survey, which is a joint project of the University of Massachusetts and the Infrared Processing and Analysis Center/California Institute of Technology, funded by the National Aeronautics and Space Administration and the National Science Foundation. 
This work used the Tool for OPerations on Catalogues And Tables (TOPCAT) processing software \citep{2005ASPC..347...29T}.   
\end{acknowledgements}

\bibliographystyle{aa}
\bibliography{references.bib}

\begin{appendix}
\section{Giant and Dwarf Sequences for VISTA Colours}
\label{appendixa}

We estimated VISTA colours of the FGKM dwarf and giant sequences by computing VISTA ZYJHK$_s$ magnitudes for 52 FGKM dwarf and 54 FGKM giant stars from the Infrared Telescope Facility (IRTF) Spectral Library\footnote{http://irtfweb.ifa.hawaii.edu/~spex/IRTF\_Spectral\_Library/}. The IRTF Spectral Library has a collection of R$\sim$2000-2500 FGKM stellar spectra observed with the SpeX spectrograph, with a spectral range of 0.8 to at least 2.5$\mu$m \citep{2005ApJ...623.1115C,2009ApJS..185..289R}. The stars in the library were absolutely flux calibrated using 2MASS photometry, with their spectral continuum shape preserved. 

The VISTA synthetic magnitude in each filter was estimated using the following equation:

\begin{equation}
\label{magnitude}
m_\lambda  = -2.5 \log_{10} \left(\int F_\lambda \, S_\lambda\,\mathrm{d}\lambda\right)  + 2.5 \log_{10}\left(\int F^{0}_\lambda \, S_\lambda\,\mathrm{d}\lambda\right). 
\end{equation}
The second term of the Equation \ref{magnitude} corresponds to the zero point of the magnitude scale. For calculation of the VISTA synthetic magnitudes presented here, the $F^0_{\lambda}$ flux corresponds to that of the Vega spectrum in the CALSPEC library\footnote{http://www.stsci.edu/hst/observatory/crds/calspec.html} \citep{2014arXiv1403.6861B}. The response function of the VISTA filter set, $S_{\lambda}$, was calculated from the quantum efficiency curve of the detector and transmission curves for each filter\footnote{http://www.eso.org/sci/facilities/paranal/instruments/vircam/inst.html}. The VISTA synthetic colours calculated for dwarf and giant stars in the IRTF library are given in Tables \ref{table_vista_colours_dwarfs} and \ref{table_vista_colours_giants}, respectively. 

Using the stars in Tables \ref{table_vista_colours_dwarfs} and \ref{table_vista_colours_giants}, we fitted the colours of the giant and dwarf sequences using a fifth-order polynomial in $H-K_s$ with the form:

\begin{equation}
\label{magnitude2}
colour \, X  = \sum\limits_{i=0}^5 A_i \left(H-K_s\right)^i.
\end{equation}

We tabulated the fitted colours of the giant and dwarf sequences as function of $H-K_s$ colour in Table \ref{table_vista_colours}. The positions of the stellar sequences are shown as orange lines in Figure \ref{colours_diag}.

\begin{table*}
\caption{VISTA synthetic magnitudes for dwarfs in the IRTF library}
\centering
\begin{tabular}{l l l l l l l}
\hline\hline
Name  &            Sp. T.  &  $Z$ &   $Y$  &  $J$ &  $H$   &   $K_s$       \\
\hline
HD 108519		& F0	& 7.265 & 7.231 & 7.113 & 6.973 & 6.935  \\ 
HD 213135		& F1	& 5.462 & 5.427 & 5.263 & 5.092 & 5.076  \\ 
HD 113139		& F2	& 4.444 & 4.372 & 4.176 & 3.998 & 3.976  \\ 
HD 26015			& F3	& 5.355 & 5.305 & 5.156 & 5.043 & 5.057  \\ 
HD 16232			& F4	& 6.298 & 6.262 & 6.081 & 5.876 & 5.848  \\ 
HD 87822			& F4	& 5.611 & 5.559 & 5.372 & 5.194 & 5.164  \\ 
HD 27524			& F5	& 6.166 & 6.144 & 5.966 & 5.788 & 5.767  \\ 
HD 218804		& F5	& 5.266 & 5.197 & 4.983 & 4.745 & 4.694  \\ 
HD 215648		& F6	& 3.577 & 3.519 & 3.299 & 3.075 & 3.042  \\ 
HD 126660		& F7	& 3.389 & 3.349 & 3.148 & 2.936 & 2.883  \\ 
HD 27383			& F8	& 6.075 & 5.999 & 5.799 & 5.589 & 5.573  \\ 
HD 219623		& F8	& 4.963 & 4.873 & 4.642 & 4.386 & 4.342  \\ 
HD 176051		& F9	& 4.376 & 4.303 & 4.039 & 3.718 & 3.667  \\ 
HD 114710		& F9.5	& 3.427 & 3.382 & 3.181 & 2.988 & 2.996  \\ 
HD 109358		& G0	& 3.504 & 3.438 & 3.195 & 2.905 & 2.876  \\ 
HD 10307			& G1	& 4.284 & 4.205 & 3.964 & 3.685 & 3.657  \\ 
HD 20619			& G1.5	& 6.274 & 6.168 & 5.878 & 5.537 & 5.475  \\ 
HD 76151			& G2	& 5.203 & 5.115 & 4.855 & 4.555 & 4.481  \\ 
HD 10697			& G3	& 5.442 & 5.315 & 5.024 & 4.695 & 4.616  \\ 
HD 214850		& G4	& 4.835 & 4.747 & 4.464 & 4.069 & 3.972  \\ 
HD 165185		& G5	& 5.097 & 5.055 & 4.831 & 4.575 & 4.525  \\ 
HD 115617		& G6.5	& 3.591 & 3.522 & 3.281 & 2.998 & 2.985  \\ 
HD 75732			& G8	& 4.938 & 4.807 & 4.504 & 4.139 & 4.047  \\ 
HD 101501		& G8	& 4.397 & 4.276 & 3.995 & 3.665 & 3.612  \\ 
HD 145675		& K0	& 5.562 & 5.438 & 5.135 & 4.796 & 4.743  \\ 
HD 10476			& K1	& 4.269 & 4.139 & 3.809 & 3.405 & 3.318  \\ 
HD 3765			& K2	& 6.173 & 6.025 & 5.678 & 5.262 & 5.183  \\ 
HD 219134		& K3	& 4.586 & 4.358 & 3.955 & 3.444 & 3.333  \\ 
HD 45977			& K4	& 7.578 & 7.403 & 7.033 & 6.555 & 6.455  \\ 
HD 36003			& K5	& 6.199 & 5.996 & 5.588 & 5.031 & 4.915  \\ 
HD 201092		& K7	& 4.014 & 3.824 & 3.427 & 2.875 & 2.736  \\ 
HD 237903		& K7	& 6.813 & 6.562 & 6.108 & 5.499 & 5.376  \\ 
HD 19305			& M0	& 7.199 & 6.932 & 6.471 & 5.837 & 5.671  \\ 
HD 209290		& M0.5	& 6.992 & 6.676 & 6.174 & 5.544 & 5.348  \\ 
HD 42581			& M1	& 5.866 & 5.555 & 5.073 & 4.465 & 4.277  \\ 
HD 36395			& M1.5	& 5.746 & 5.407 & 4.893 & 4.215 & 4.032  \\ 
Gl 806			& M2	& 8.107 & 7.786 & 7.304 & 6.741 & 6.581  \\ 
HD 95735			& M2	& 4.928 & 4.611 & 4.125 & 3.611 & 3.435  \\ 
Gl 381			& M2.5	& 7.895 & 7.522 & 6.993 & 6.438 & 6.236  \\ 
Gl 581			& M2.5	& 7.545 & 7.163 & 6.649 & 6.123 & 5.887  \\ 
Gl 273			& M3.5	& 6.679 & 6.225 & 5.668 & 5.149 & 4.915  \\ 
Gl 388			& M3	& 6.334 & 5.936 & 5.403 & 4.841 & 4.645  \\ 
Gl 213			& M4	& 8.141 & 7.665 & 7.112 & 6.641 & 6.415  \\ 
Gl 299			& M4	& 9.392 & 8.933 & 8.377 & 7.958 & 7.711  \\ 
Gl 51			& M5	& 9.799 & 9.225 & 8.575 & 8.029 & 7.758  \\ 
Gl 406			& M6	& 8.613 & 7.819 & 7.054 & 6.483 & 6.131  \\ 
GJ 1111			& M6.5	& 9.761 & 8.954 & 8.181 & 7.648 & 7.295  \\ 
Gl 644C			& M7	& 11.357 & 10.533 & 9.741  & 9.224  & 8.862    \\
LP 412-31			& M8	& 13.672 & 12.681 & 11.732 & 11.089 & 10.675   \\
DENIS-P J1048-3956		& M9	& 11.456 & 10.389 & 9.459  & 8.933  & 8.537    \\
LP 944-20				& M9	& 12.933 & 11.718 & 10.664 & 10.052 & 9.605    \\
BRIB 0021-0214	 	& M9.5	& 14.195 & 12.995 & 11.843 & 11.132 & 10.625   \\
\hline
\end{tabular}
\label{table_vista_colours_dwarfs}
\end{table*}

\begin{table*}
\caption{VISTA synthetic magnitudes for giants in the IRTF library}
\centering
\begin{tabular}{l l l l l l l}
\hline\hline
Name  &            Sp. T.  &  $Z$ &   $Y$  &  $J$ &  $H$   &   $K_s$       \\
\hline
HD 89025	& F0	&	2.956 	& 2.914 	&  2.763 	&  2.606 	&  2.578   \\
HD 21770	& F4	&	4.683 	& 4.645 	&  4.473 	&  4.295 	&  4.275   \\
HD 17918	& F5	&	5.671 	& 5.599 	&  5.412 	&  5.224 	&  5.215   \\
HD 124850	& F7	&	3.431 	& 3.352 	&  3.121 	&  2.883 	&  2.862   \\
HD 220657	& F8	&	3.913 	& 3.809 	&  3.521 	&  3.196 	&  3.093   \\
HD 6903		& F9	&	4.925 	& 4.813 	&  4.535 	&  4.216 	&  4.155   \\
HD 21018	& G1	& 	5.305 	& 5.146 	&   4.827 	&  4.475 	&  4.391   \\
HD 88639	& G3	&	4.988 	& 4.837 	&  4.504 	&  4.111 	&  4.049   \\
HD 108477	& G4	&	5.299 	& 5.135 	&  4.799 	&  4.402 	&  4.312   \\
HD 193896	& G5	&	5.305 	& 5.095 	&  4.712 	&  4.265 	&  4.151   \\
HD 27277	& G6	&	6.808 	& 6.587 	&  6.215 	&  5.795 	&  5.715   \\
HD 182694	& G7	&	4.865 	& 4.729 	&  4.395 	&  3.988 	&  3.888   \\
HD 16139	& G7.5	&	6.755 	& 6.549 	&  6.173 	&  5.715 	&  5.595   \\
HD 135722	& G8	&	2.174 	& 2.006 	&  1.645 	&  1.153 	&  1.034   \\
HD 104979	& G8	&	3.108 	& 2.944 	&  2.589 	&  2.105 	&  2.013   \\
HD 122563	& G8	&	4.969 	& 4.785 	&  4.392 	&  3.862 	&  3.765   \\
HD 222093	& G9	&	4.692 	& 4.508 	&  4.123 	&  3.619 	&  3.517   \\
HD 100006	& K0	&	4.511 	& 4.295 	&  3.893 	&  3.363 	&  3.261   \\
HD 9852		& K0.5	&	6.183 	& 5.859 	&  5.356 	&  4.755 	&  4.605   \\
HD 25975	& K1	&	4.979 	& 4.849 	&  4.522 	&  4.098 	&  4.028   \\
HD 91810	& K1	&	5.497	& 5.256 	&  4.831 	&  4.303 	&  4.169   \\
HD 36134	& K1	&	4.415	& 4.209 	&  3.797 	&  3.259 	&  3.165   \\
HD 124897	& K1.5	&	-1.535	&-1.759 	&  -2.203 	& -2.853 	& -2.987  \\
HD 137759	& K2	&	1.998 	& 1.756 	&  1.314 	&  0.747 	&  0.611   \\
HD 132935	& K2	&	5.206 	& 4.931 	&  4.441 	&  3.779 	&  3.614   \\
HD 2901		& K2	&	5.653 	& 5.391 	&  4.931 	&  4.325 	&  4.225   \\
HD 99998	& K3	&	3.037 	& 2.717 	&  2.159 	&  1.414 	&  1.215   \\
HD 35620	& K3	&	3.637 	& 3.336 	&  2.845 	&  2.227 	&  2.048   \\
HD 178208	& K3	&	5.185 	& 4.892 	&  4.411 	&  3.821 	&  3.674   \\
HD 221246	& K3	&	4.675 	& 4.364 	&  3.863 	&  3.203 	&  3.055   \\
HD 114960	& K3.5	&	4.852 	& 4.558 	&  4.088 	&  3.471 	&  3.316   \\
HD 207991	& K4	&	4.909 	& 4.548 	&  3.983 	&  3.209 	&  3.005   \\
HD 181596	& K5	&	5.721 	& 5.357 	&  4.799 	&  4.073 	&  3.867   \\
HD 120477	& K5.5	&	2.172 	& 1.823 	&  1.268 	&  0.532 	&  0.316   \\
HD 3346		& K6	&	3.126 	& 2.786 	&  2.225 	&  1.452 	&  1.213   \\
HD 194193	& K7	&	4.025 	& 3.646 	&  3.064 	&  2.252 	&  2.022   \\
HD 213893	& M0	&	4.812 	& 4.444 	&  3.885 	&  3.125 	&  2.916   \\
HD 204724	& M1	&	2.488 	& 2.116 	&  1.593 	&  0.875 	&  0.618   \\
HD 120052	& M2	&	2.894 	& 2.488 	&  1.864 	&  1.022 	&  0.747   \\
HD 219734	& M2.5	&	2.519 	& 2.098 	&  1.493 	&  0.681 	&  0.428   \\
HD 39045	& M3	&	3.621 	& 3.181 	&  2.551 	&  1.719 	&  1.446   \\
HD 28487	& M3.5	&	3.851 	& 3.324 	&  2.659 	&  1.781 	&  1.485   \\
HD 27598	& M4	&	4.038 	& 3.562 	&  2.953 	&  2.128 	&  1.856   \\
HD 214665	& M4	&	2.556 	& 1.886 	&  1.114 	&  0.156 	& -0.182  \\
HD 4408		& M4	&	2.539 	& 2.043 	&  1.405 	&  0.546 	&  0.239   \\
HD 204585	& M4.5	&	2.294 	& 1.691 	&  1.062 	&  0.278 	&  0.026   \\
HD 175865	& M5	&	0.496 	& -0.128   	& -0.769 	& -1.570 	& -1.811   \\
HD 94705	& M5.5	&	1.822 	& 1.089 	&  0.401 	& -0.429 	& -0.763   \\
HD 18191	& M6	&	1.739 	& 0.966 	&  0.252 	& -0.622 	& -0.938    \\
HD 196610	& M6	&	1.653 	& 0.816 	&  0.144 	& -0.649 	& -0.945    \\
HD 108849	& M7	&	2.488 	& 1.362 	&  0.495 	& -0.332 	& -0.722    \\
HD 207076	& M7	&	1.186 	& 0.195 	&  -0.500 	& -1.206 	& -1.477   \\
IRAS 21284-0747	& M8	&	8.752 	& 7.262 	&  6.107 	&  5.373 	&  4.819    \\
BRIB 1219-1336	& M9	&	10.689	& 9.446 	&  8.556 	&  7.919 	&  7.447    \\
\hline
\end{tabular}
\label{table_vista_colours_giants}
\end{table*}

\begin{table*}
\caption{VISTA colours as function of $(H-K_s)$ for dwarfs and giants}
\centering
\begin{tabular}{@{\extracolsep{4pt}}l l l l l l l@{}}
\hline\hline
\multicolumn{3}{c}{Dwarfs} & \multicolumn{3}{c}{Giants} \\
\cline{1-3}\cline{4-6} 
 $(Z-Y)$ &   $(Y-J)$  &  $(J-H)$ &    $(Z-Y)$ &   $(Y-J)$  &  $(J-H)$ & $(H-K_s)$       \\
\hline
     0.065  & 0.181   & 0.169     &	 &	   &	         &-0.015\\
    0.052   & 0.176   & 0.162     &	 &	   &	         &-0.005\\
    0.046   & 0.178   & 0.168     & 0.027   & 0.146   & 0.126   & 0.005\\
    0.046   & 0.186   & 0.183     & 0.051   & 0.177   & 0.176   & 0.015\\
    0.051   & 0.198   & 0.205     & 0.072   & 0.207   & 0.221   & 0.025\\
    0.059   & 0.214   & 0.234     & 0.091   & 0.234   & 0.263   & 0.035\\
    0.070   & 0.232   & 0.266     & 0.109   & 0.260   & 0.301   & 0.045\\
    0.084   & 0.253   & 0.300     & 0.125   & 0.284   & 0.337   & 0.055\\
    0.100   & 0.274   & 0.335     & 0.140   & 0.307   & 0.370   & 0.065\\
    0.117   & 0.296   & 0.370     & 0.154   & 0.328   & 0.401   & 0.075\\
    0.136   & 0.318   & 0.405     & 0.168   & 0.348   & 0.431   & 0.085\\
    0.155   & 0.340   & 0.437     & 0.182   & 0.368   & 0.460   & 0.095\\
    0.174   & 0.362   & 0.467     & 0.195   & 0.386   & 0.487   & 0.105\\
    0.194   & 0.383   & 0.494     & 0.209   & 0.404   & 0.514   & 0.115\\
    0.214   & 0.403   & 0.517     & 0.223   & 0.421   & 0.540   & 0.125\\
    0.233   & 0.422   & 0.537     & 0.238   & 0.438   & 0.565   & 0.135\\
    0.253   & 0.439   & 0.553     & 0.253   & 0.454   & 0.589   & 0.145\\
    0.273   & 0.456   & 0.565     & 0.269   & 0.470   & 0.613   & 0.155\\
    0.293   & 0.472   & 0.573     & 0.286   & 0.486   & 0.637   & 0.165\\
    0.313   & 0.486   & 0.578     & 0.304   & 0.501   & 0.659   & 0.175\\
    0.333   & 0.500   & 0.580     & 0.324   & 0.516   & 0.681   & 0.185\\
    0.354   & 0.513   & 0.578     & 0.344   & 0.531   & 0.703   & 0.195\\
    0.374   & 0.525   & 0.574     & 0.366   & 0.545   & 0.723   & 0.205\\
    0.396   & 0.537   & 0.567     & 0.389   & 0.559   & 0.742   & 0.215\\
    0.418   & 0.549   & 0.559     & 0.414   & 0.574   & 0.761   & 0.225\\
    0.441   & 0.560   & 0.549     & 0.439   & 0.588   & 0.778   & 0.235\\
    0.465   & 0.572   & 0.539     & 0.467   & 0.602   & 0.794   & 0.245\\
    0.490   & 0.585   & 0.528     & 0.495   & 0.615   & 0.808   & 0.255\\
    0.516   & 0.598   & 0.517     & 0.525   & 0.629   & 0.820   & 0.265\\
    0.544   & 0.612   & 0.507     & 0.556   & 0.643   & 0.831   & 0.275\\
    0.574   & 0.628   & 0.498     & 0.589   & 0.656   & 0.840   & 0.285\\
    0.605   & 0.644   & 0.491     & 0.623   & 0.670   & 0.847   & 0.295\\
    0.637   & 0.662   & 0.485     & 0.658   & 0.683   & 0.852   & 0.305\\
    0.672   & 0.682   & 0.482     & 0.693   & 0.697   & 0.854   & 0.315\\
    0.707   & 0.703   & 0.482     & 0.730   & 0.710   & 0.855   & 0.325\\
    0.745   & 0.726   & 0.484     & 0.768   & 0.723   & 0.853   & 0.335\\
    0.783   & 0.751   & 0.490     & 0.807   & 0.737   & 0.849   & 0.345\\
    0.823   & 0.778   & 0.498     & 0.846   & 0.750   & 0.843   & 0.355\\
    0.864   & 0.806   & 0.510     & 0.885   & 0.763   & 0.834   & 0.365\\
    0.906   & 0.836   & 0.525     & 0.925   & 0.777   & 0.824   & 0.375\\
    0.947   & 0.867   & 0.543     & 0.965   & 0.791   & 0.811   & 0.385\\
    0.989   & 0.899   & 0.563     & 1.005   & 0.805   & 0.798   & 0.395\\
    1.030   & 0.932   & 0.585     & 1.045   & 0.819   & 0.782   & 0.405\\
    1.069   & 0.964   & 0.609     & 1.085   & 0.834   & 0.766   & 0.415\\
    1.106   & 0.997   & 0.633     & 1.124   & 0.849   & 0.748   & 0.425\\
    1.141   & 1.028   & 0.657     & 1.162   & 0.865   & 0.731   & 0.435\\
    1.171   & 1.058   & 0.679     & 1.199   & 0.881   & 0.713   & 0.445\\
    1.197   & 1.086   & 0.699     & 1.236   & 0.898   & 0.696   & 0.455\\
    1.217   & 1.110   & 0.714     & 1.270   & 0.916   & 0.680   & 0.465\\
    1.230   & 1.129   & 0.724     & 1.304   & 0.935   & 0.666   & 0.475\\
    1.234   & 1.144   & 0.726     & 1.335   & 0.956   & 0.655   & 0.485\\
    1.229   & 1.151   & 0.718     & 1.364   & 0.978   & 0.647   & 0.495\\
    1.212   & 1.150   & 0.698     & 1.391   & 1.001   & 0.644   & 0.505\\
    1.182   & 1.140   & 0.663     & 1.416   & 1.026   & 0.646   & 0.515\\
   	     &	       &	   & 1.437   & 1.054   & 0.654   & 0.525\\
   	     &	       &	   & 1.456   & 1.084   & 0.670   & 0.535\\
   	     &	       &	   & 1.471   & 1.116   & 0.695   & 0.545\\
   	     &	       &	   & 1.482   & 1.151   & 0.731   & 0.555\\
\hline
\end{tabular}
\label{table_vista_colours}
\end{table*}

\end{appendix}

\end{document}